\documentstyle[prl,aps]{revtex}

\def\C{{\cal C}}
\def\A{{\cal A}}
\def\E{{\cal E}}
\def\H{{\cal E}}
\def\K5{\kappa_5}

\begin{document}
\preprint{hep-th/0006007 v2}
\draft

\twocolumn[\hsize\textwidth\columnwidth\hsize\csname
@twocolumnfalse\endcsname

\title{Gravitational waves from inflation on the brane}

\author{David Langlois$^1$, Roy Maartens$^2$ and David Wands$^2$}
\address{$^1$Departement d'Astrophysique Relativiste et de Cosmologie,
 Observatoire de Paris, 92195~Paris, France}
\address{$^2$Relativity and Cosmology Group,
School of Computer Science and Mathematics,\\ University of
Portsmouth, Portsmouth~PO1~2EG, Britain}

\date{\today}

\maketitle

\begin{abstract}

We discuss the evolution of gravitational waves in a brane-world
cosmology embedded in five-dimensional anti-de Sitter spacetime.
We show that during slow-roll inflation, modelled as a period of
quasi-de Sitter expansion on the brane, there is a discrete
normalizable massless graviton mode. There is a mass gap due to the
expansion, above which there is a continuum of massive modes. Only
the massless mode is `light' compared with the Hubble scale during
inflation, leading to the production of classical perturbations on
large scales from vacuum fluctuations on small scales. We
calculate the amplitude of these fluctuations at horizon-crossing
and show that the standard four-dimensional result is recovered at
low energies, but the amplitude of the perturbations is enhanced
at high energies.

\end{abstract}

\pacs{~ \hfill hep-th/0006007 v2}

\vskip2pc]

\section{Introduction}

The idea that there may exist large extra dimensions, inspired by
recent developments in M/string-theory, has recently attracted much
attention. This possibility arises when one assumes that ordinary
matter is confined to a brane, our `Universe', within a
higher-dimensional spacetime. In this spirit, Randall and Sundrum
\cite{RS99} proposed a five-dimensional model with a curved metric,
for which they showed that the linearized gravitational interaction
behaves effectively as standard four-dimensional gravity (plus
correction terms), even though the fifth dimension may be large,
possibly infinite.

Although appealing, these models have still to be confronted with
observations. One line of investigation is to search for exotic
effects in accelerators. Another, to which the present work belongs,
is to explore the consequences of these models in cosmology. The first
step, undertaken during the last year, has been to study the
homogeneous and isotropic cosmologies of brane models, and to see
whether they are compatible with cosmological evolution.  The next
step is to study the cosmological perturbations and confront them with
the increasingly detailed observations of large scale structure and
cosmic microwave background (CMB) anisotropies.

Cosmological perturbations in brane models are however more
complicated to deal with than in the standard theory of cosmological
perturbations. The general formalism has only very recently been
investigated~\cite{KIS,M,Langlois,BDBL,KS}. In the present work, our
goal is to study the evolution of gravitational waves, which are not
related directly to the matter perturbations in the brane, and are
thus in principle easier to study. However, the equation of motion for
gravitational waves is complicated because in general one must solve a
two-variable partial differential equation, even after decomposition
into Fourier modes. In the Randall-Sundrum (RS) scenario of a
Minkowski background brane, the equation separates and may be solved
exactly~\cite{RS99}. For an expanding background brane, this is in
general no longer possible in coordinates in which the brane is fixed
in the bulk. It turns out, however, that there is one case where it is
possible: de Sitter evolution on the brane. Thus we are able to make
quantitative predictions about the production and evolution of
gravitational waves during slow-roll inflation on the brane. Scalar
perturbations in this scenario have been previously investigated
in~\cite{MWBH99}.  As in the non-expanding solutions, the massless
mode is decoupled from the massive modes and it is possible to avoid
contamination of the massless graviton modes by massive modes on the
brane during slow-roll inflation. If this were not the case, there
would be potentially problematic production of massive graviton modes
during inflation, as pointed out in~\cite{KIS}.

\section{Cosmological tensor perturbations}

\subsection{Background metric}

Before addressing the cosmological perturbations, let us specify
the background five-dimensional spacetime we are going to use. The
metric is of the form
\begin{equation}
g_{AB}dx^A dx^B
= -n^{2}(t,y) dt^2+a^{2}(t,y)
d\vec{x}\,^2+dy^{2}\,, \label{metric}
\end{equation}
where $x^A=(x^\mu,x^5)=(t,x^i,y)$. In the following, a dot will
denote $\partial/\partial t$ and a prime $\partial /\partial y$.
Note that we have assumed, for simplicity, that the spatial
three-surfaces are flat. Cosmological solutions can then be
obtained~\cite{BDL} by solving the five-dimensional Einstein
equations,
\begin{equation}
G_{AB}\equiv R_{AB}-{\textstyle{1\over 2}} R g_{AB}=\K5^2
T_{AB}\,, \label{einstein}
\end{equation}
where $R_{AB}$ is the five-dimensional Ricci tensor and
$R=g^{AB}R_{AB}$ its trace. The gravitational constant is related
to the five-dimensional Planck scale, $\kappa_5^2\equiv
8\pi/M_5^3$. The energy-momentum tensor on the right hand side can
be decomposed into
\begin{equation}
{T}^A{}_B  =- \K5^{-2} \Lambda_5\delta^A{}_B+ \delta (y)S^A{}_{B}\,,
\end{equation}
where we assume there is only a cosmological constant in the bulk,
thus generalizing the RS scenario to cosmology. The term $S^A{}_B$
is the energy-momentum tensor of the matter localized on the brane
$(y=0)$. Because of the symmetries of the metric in
Eq.~(\ref{metric}), this is necessarily of perfect-fluid form:
\begin{equation}
S^A{}_{B}=  \mbox{diag} \left(-\rho_{\rm b},p_{\rm b},p_{\rm
b},p_{\rm b},0 \right), \label{source}
\end{equation}
where the energy density $\rho_{\rm b}$ and pressure $p_{\rm b}$
are functions only of time $t$. Also, because $T^0_{\, 5}=0$, it follows 
from Einstein's equations that 
\begin{equation} 
\left({\dot a \over n}\right)'=0,
\end{equation}
which implies that, if one chooses $t$ to coincide with the  proper time 
on the brane, one gets 
\begin{equation}
n={\dot a\over\dot{a}_o}\,.\label{na1}
\end{equation}

The influence of the brane matter
on the five-dimensional metric enters via the junction conditions,
\begin{equation}
\left[K_{\mu\nu}\right]=-\K5^2\left(S_{\mu\nu}-{\textstyle{1\over
3}} S g_{\mu\nu}\right)\,, \label{junction}
\end{equation}
where the square brackets denote the `jump' across the brane
($[f]\equiv f(0^+)-f(0^-)$), and $K_{\mu\nu}$ is the extrinsic
curvature tensor of the brane. This implies for the metric
components~\cite{BDL}
\begin{equation}
{[a']\over a_o}=-{\K5^2\over 3}\rho_{\rm b}\,, \qquad {[n']\over
n_o}={\K5^2\over 3}\left(3p_{\rm b}+2\rho_{\rm b}\right)\,,
\label{flrwjunction}
\end{equation}
where the subscript `o' denotes the value taken on $y=0$. Assuming
in addition $Z_2$ symmetry, the jump is then simply twice the
value on one side.

It is then possible to find, using  Einstein's
equations~(\ref{einstein}), a first integral that is analogous to
the Friedmann equation, but which reads~\cite{BDEL} (see also
\cite{SchAdS,SMS})
\begin{equation}
H_o^2\equiv {\dot a_o^2\over a_o^2}={ \Lambda_5 \over 6}
+ { \kappa_5^4 \over 36}\rho_{\rm b}^2+{\C \over a_o^4}\,,
\label{newfriedmann}
\end{equation}
where $\C$ is  an integration constant which can be interpreted as
arising from a five-dimensional tidal effect (see~\cite{SMS,M}).
Combining Eq.~(\ref{newfriedmann}) with the conservation equation
\begin{equation}
\dot{\rho}_{\rm b}+3{\dot{a}_o\over a_o}(\rho_{\rm b}+p_{\rm
b})=0\, \label{cons}
\end{equation}
implied by Einstein's equations~(\ref{einstein}), which is the
same as in standard cosmology, one can obtain the cosmological
evolution in the brane.

In general, the scale factor $a_o(t)$, does not follow the
conventional  cosmological evolution~\cite{BDL}. However, standard
cosmology can be recovered at low energies~\cite{cosmors,BDEL} in
the particular case where the brane is endowed with a constant
tension $\lambda$, so that
\[
\rho_{\rm b}=\rho+\lambda\,,~~p_{\rm b}=p-\lambda\,.
\]
This allows one (when $\C=0$) to regain the general relativistic
Friedmann equation from Eq.~(\ref{newfriedmann}) in the low-energy
limit $\rho\ll\lambda$, by identifying
\begin{equation}
\kappa_4^2 \equiv {\textstyle{1\over6}}\K5^4\lambda \,,
\end{equation}
where the effective four-dimensional Planck mass is given by
$M_4^2=8\pi/\kappa_4^2$. Then Eq.~(\ref{newfriedmann}) becomes
\begin{equation}
H_o^2 = {\Lambda_4\over3} +
{\kappa_4^2\over3}\rho\left(1+{\rho\over 2\lambda}\right)+{\C
\over a_o^4} \,,\label{newf}
\end{equation}
where $\Lambda_4=(\Lambda_5+{1\over6}\K5^4\lambda^2)/2$.
If the brane tension $\lambda$ is adjusted so as to compensate for
the (negative) bulk cosmological constant $\Lambda_5$ (so that
$\Lambda_4=0$), we require
\[
\K5^2\lambda = {\sqrt{-6\Lambda_5}}\,.
\]

Finally, one can solve for the metric in the bulk, to get the explicit
dependence of the metric on the coordinates $y$ and $t$~\cite{BDEL}.
The bulk metric can be shown to be a Schwarzschild-anti de Sitter
solution, where the Schwarzschild mass parameter is proportional to
$\C$~\cite{MSM}. For solutions with $\C=0$, it follows that the bulk
is anti-de Sitter, and thus conformally flat, while $\C\neq0$ signals
a non-conformally flat bulk.

\subsection{Tensor perturbations}

Linear perturbations in brane cosmology have been studied very
recently~\cite{KIS,M,Langlois,BDBL,KS}. In the present work, we
will focus only on gravitational waves (or tensor modes in the
usual terminology in cosmology~\cite{MFB}), and leave aside the
scalar and vector modes. Gravitational waves can propagate
independently of the matter content in the universe, and they are
therefore simpler to study. Moreover, these modes were the first
to be investigated in the non-cosmological (static)
case~\cite{RS99}.

Following~\cite{Langlois}, we will thus consider a perturbed
metric of the form
\begin{equation}
\label{pertds}
ds^2=-n^2 dt^2+a^2\left(\delta_{ij}+E_{ij}\right)dx^i dx^j+dy^2\,,
\end{equation}
where $E_{ij}$ is transverse traceless, i.e., $\partial_iE^{ij}=0$
and $E^i{}_i=0$ (using the metric $\delta_{ij}$ to raise or lower
the spatial indices). Note that these tensor perturbations are
automatically gauge-invariant, i.e. are not affected by a
coordinate change to first order.

The equation of motion for the tensor modes is then obtained from
the linear perturbations of Einstein's equations. In the bulk,
since there is only a cosmological constant, these reduce
simply to
\begin{equation}
\delta R_{ij}={\textstyle{2\over 3}}\Lambda_5 E_{ij}\,,
\end{equation}
which yields the following equation of motion~\cite{Langlois}:
\begin{eqnarray}
&& {1\over n^{2}}\ddot{E} + {1\over n^2} \left({3\dot a\over
a}-{\dot n\over n} \right)\dot E+{k^2\over a^2}E\nonumber\\ &&
-E'' - \left(3{a'\over a}+{n'\over n}\right)E' = 0\,,
\label{eqmotion}\label{eqm2}
\end{eqnarray}
where we have decomposed the general metric perturbation into
Fourier modes:
\[
E^i{}_j(t,\vec{x},y) = E(t,y;\vec{k})\, \exp({\rm
i}\vec{k}\cdot\vec{x})\, 
\hat{e}^i{}_j\,,
\]
with $\hat{e}_{ij}$ a transverse traceless polarization tensor.
We can rewrite Eq.~(\ref{eqmotion}) as
\[
\left({a^3\over n}\dot{E}
\right)^{\!\!\displaystyle{\cdot}}+k^2anE-\left(a^3n E'
\right)'=0\,.
\]
These equations simplify dramatically if $n=a$ and $\dot n=0=\dot a$,
which is the RS case, for which the equation separates and may be
solved exactly~\cite{RS99}. For an expanding background brane,
separation is in general no longer possible in coordinates in which
the brane remains fixed in the bulk. It is always possible (for
$\C=0$) to write the general solution for linear perturbations about
five-dimensional anti-de Sitter spacetime in a manifestly static
coordinate system~\cite{KS}, but it is non-trivial to impose the
boundary conditions at the brane, which is then moving.

The wave equation~(\ref{eqm2}) is to be solved using the background
form of $a(t,y)$, as given in Ref. \cite{BDEL}. The boundary condition
at the brane follows from the linearly perturbed junction conditions
given in Eq.~(\ref{junction}):
\[
E'_{ij}\Big|_{y=0^+}= \bar \pi_{ij}\,,
\]
where $\bar \pi_{ij}$ is the tensor contribution of the anisotropic
stress exerted by matter on the brane, which we will henceforth assume
to be zero. Thus we set
\begin{equation}
E'|_o= 0\,. \label{boundary}
\end{equation}
However, we note that $E''$ in Eq.~(\ref{eqm2}) remains nonzero on
the brane, and we need to solve for the $y$-dependence of $E$.

In order to gain further insight, we can expand in the off-brane
direction. Using the boundary conditions in Eqs.~(\ref{flrwjunction})
and (\ref{boundary}) and taking $t$ as the brane proper time, we find
that
\begin{eqnarray*}
a(t,y) &=&
a_o(t)\left[1-{\textstyle{1\over 6}}\K5^2\{\rho(t)+\lambda\}|y|+
O(y^2)\right]\,,\\ 
n(t,y) &=& 1+{\textstyle{1\over 6}}\K5^2\{3p(t)+2\rho(t)-\lambda\}|y|+
O(y^2)\,,\\
E(t,y) &=&
E_o(t)+{\textstyle{1\over2}}E_2(t)y^2+
{\textstyle{1\over6}}E_3(t)|y|^3+ O(y^4)\,,
\end{eqnarray*}
where $E_n(t)=\partial^n E(t,0^+)/\partial y^n$. Then expanding 
Eq.~(\ref{eqm2}) order by order yields
\begin{eqnarray}
 E_2 &=& \ddot{E}_o+3H_o\dot{E}_o+{k^2\over a_o^2}E_o\,,\label{e2}\\
E_3&=& -{\textstyle{1\over2}}\K5^2(\rho+3p-2\lambda)E_2
\nonumber\\ &&~{} +{\textstyle{1\over2}}
\K5^2(\rho+p)\left[(5+3c_{\rm s}^2)H_o\dot{E}_o+2{k^2\over
a_o^2}E_o\right]\,,\label{e3}\\ 
(\cdots) &&  \,,\nonumber
\end{eqnarray}
where $c_{\rm s}^2=\dot{p}/\dot{\rho}$. These equations determine
the $y$-derivatives on the brane of the tensor perturbations, in
terms of $E_o$. The 5-dimensional bulk equation~(\ref{eqm2}) is
replaced by the infinite sequence of equations~(\ref{e2}),
(\ref{e3}), (\dots), on the brane.

If $E_2=0$, i.e., if $E''=0$ on the brane, then Eq.~(\ref{e2}) reduces
to the standard general relativity equation for tensor perturbations,
generalizing the RS zero-mode~\cite{RS99} to a cosmological
context. Its solution $E_o(t)$ then fixes $E_n$, for $n\geq3$, via the
subsequent equations. This shows that $E=E_o(t)$, where $E_n=0$ for
all $n\geq1$, is inconsistent in general, since $E_o$ is subject to an
infinite chain of constraints. In general the ``zero-mode" is coupled
to the higher-order terms.

However there is one important limit in which the ``zero-mode" remains
decoupled. On large scales, where we can neglect the $k^2$ term, the
gravitational wave equation~(\ref{eqm2}) always admits the homogeneous
solution $E(t,y;k\to0)=E_o=$~constant, which, of course, satisfies the
boundary condition in Eq.~(\ref{boundary}).  Thus the standard result
from general relativity in four dimensions that the amplitude of
tensor perturbations stays constant on super-horizon scales remains
true in a brane-world cosmology\footnote{A similar result can be
obtained for scalar metric perturbations corresponding to adiabatic
density perturbations on large scales~\cite{MWBH99,WMLL00}.}.

The other case in which terms involving $E_0$ drop out of Eq. (\ref{e3}) 
for $E_3$ and all higher $E_n$ is when $\rho+p=0$, corresponding 
to de Sitter expansion in the brane,  and we will now 
consider this case in greater detail.

\section{Bulk gravitons with a de Sitter brane}

\subsection{Separable background}

In order  to solve for the $y$-dependence of the bulk
gravitons and to study the time-dependence of the perturbations on
the brane, we will consider separable solutions for the bulk
metric, i.e., where the scale factor is of the form
\begin{equation}
a(t,y)=a_o(t)\A(y) \,,~~\A(0)=1\,. \label{asep}
\end{equation}
It follows from Eq.~(\ref{na1}) that
\begin{equation}
n=\A(y) \,, \label{n}
\end{equation}
and then the junction conditions in Eq.~(\ref{flrwjunction})
immediately require that the stresses on the brane obey the
equation of state $p_{\rm b}=-\rho_{\rm b}=$constant, so that
$p=-\rho=$ constant. This is the case for a constant scalar field
on the brane, and it will be a good approximation to a
potential-dominated scalar field rolling slowly down a
sufficiently flat potential~\cite{MWBH99}.

Substituting (\ref{asep}-\ref{n}) into Einstein's equations 
shows that $\C=0$, so that the bulk is
conformally flat anti de Sitter spacetime. The generalized
Friedmann equation~(\ref{newf}) then shows that $H_o=$ constant, so
that $a_o(t)=\exp(H_ot)$. Thus a separable scale factor
in the coordinates of Eq.~(\ref{metric}) arises if and only if the
induced metric on the brane is de Sitter (including the RS static
case). Solving for the metric in the bulk then yields~\cite{bent}
\begin{equation}
\label{Ay}
\A(y)= \cosh(\mu y)-\left[1+{\rho\over\lambda}\right]\sinh(\mu
|y|)\,,
\end{equation}
where
\begin{equation}
\mu = {\kappa_4^2\over\K5^2}={M_5^3\over M_4^2}\,.
\end{equation}
Constant-$y$ hypersurfaces correspond to exponentially expanding de
Sitter slices for $\rho>0$, giving a dS$_4$ slicing of AdS$_5$. The
original RS solution with Minkowski spacetime on the brane (M$_4$
slicing of AdS$_5$) is recovered in the limit $\rho/\lambda\to0$, when
$\A\to \exp(-\mu|y|)$.  At very high energies, $\rho\gg\lambda$,
deviations from the RS solution will be significant.

The $y$-dependence of the scale factor can be conveniently
rewritten in the form
\begin{equation}
\A(y) = {H_o\over\mu}\sinh\mu(y_{\rm h}- |y|)\,,
\end{equation}
where $y=\pm y_{\rm h}$ are Cauchy horizons ($g_{00}(\pm y_{\rm
h}) =0$)~\cite{bent}, with
\begin{equation}
y_{\rm h} ={1\over \mu}\coth^{-1}\left(1+{\rho\over\lambda}\right)
\,.
\end{equation}
In the RS model, $y_{\rm h} \to\infty$. It is useful to introduce
the conformal bulk-coordinate $z=\int dy/\A(y)$:
\begin{equation}
z = {\rm sgn}(y) H_o^{-1}
\ln\left[\coth{\textstyle{1\over2}}\mu(y_{\rm h}-|y|) \right]\,.
\label{defz}
\end{equation}
The Cauchy horizon is now at $|z|=\infty$, and the brane is
located at $z=\pm z_{\rm b}$, with
\begin{equation}
z_{\rm b} ={1\over  H_o}\sinh^{-1}{H_o\over\mu}\,.
\end{equation}
The line element, Eq.~(\ref{metric}), becomes
\begin{equation}
ds^{2} = \A^2(z) \left[ -dt^2 + dz^2 +{\rm e}^{2H_ot}d\vec{x}\,^2
\right]\,, \label{dSmetric}
\end{equation}
where
\begin{equation}
\A(z) ={ H_o\over \mu \sinh(H_o|z|)}\,. \label{A}
\end{equation}
The $(t,z)$ coordinates are well-suited to describing the causal
structure of the bulk inside the Cauchy horizon, which is
conformal to Minkowski spacetime with the region $|z|<z_{\rm b}$
cut out and the $z=\pm z_{\rm b}$ hypersurfaces identified. (The
brane tension generates the discontinuity in the extrinsic
curvature either side of $z=\pm z_{\rm b}$.) In particular this
will enable us to treat the bulk gravitons as perturbations
evolving in a 2-D Minkowski spacetime.

\subsection{Separation of perturbation variables}

In the dS$_4$ slicing of AdS$_5$, the gravitational wave
equation~(\ref{eqm2}) reduces to
\begin{equation}
\ddot{E}+3H_o\dot{E}+{k^2\over a_o^2}E=\A^2 E''+4\A \A'E'\,,
\label{E}
\end{equation}
which can then be separated into eigenmodes of the time-dependent
equation on the brane and the off-brane equation:
\[
E(t,y;\vec{k})= \int dm\, \varphi_m(t;\vec{k})\H_m(y) \,,
\]
where
\begin{eqnarray}
\ddot{\varphi}_m +3H_o\dot{\varphi}_m+\left[ m^2+{k^2\over
a_o^2}\right] \varphi_m &=&0\,, \label{varphieom}\\
\H_m''+4{\A'\over\A}\H_m'+ {m^2\over \A^2}\H_m &=& 0\,.
\label{Heom}
\end{eqnarray}
We recover the RS solutions in the limit $H_o\to0$, $\rho\to0$,
in which case $\varphi_m=\exp(\pm {\rm i}\omega t)$, with
$\omega^2=k^2+m^2$, and $\H_m$ can be given in terms of Bessel
functions of order 2~\cite{RS99}.

For $H_o>0$, the time-dependent function $\varphi_m(t;\vec{k})$
obeys the wave equation~(\ref{varphieom}) for a massive scalar
field in de Sitter 4-spacetime. If we write $u_m=a_o\varphi_m$ and
work in terms of the conformal time $\eta=-1/(a_oH_o)$, the wave
equation can be rewritten as
\[
{d^2 u_m \over d\eta^2}
 + \left[ k^2 - {2-(m^2/H_o^2) \over \eta^2} \right] u_m = 0 \,,
\]
whose general solution is given by
\[
u_m(\eta;\vec{k}) = \sqrt{-k\eta}\, B_\nu(-k\eta)\,, ~~
\nu^2={9\over4}-{m^2\over H_o^2}\,,
\]
where $B_\nu$ is a linear combination of Bessel functions of order
$\nu$. The solutions oscillate at early-times/small-scales for all
$m$, with an approximately constant amplitude while they remain
within the de Sitter event horizon ($k\gg a_oH_o$).  `Heavy
modes', with $m>{3\over2}H_o$, continue to oscillate as they are
stretched to super-horizon scales, but their amplitude rapidly
decays away, $|u_m^2|\propto a_o^{-3}$. But for `light modes' with
$m<{3\over2}H_o$, the perturbations become over-damped at
late-times/large-scales ($k\ll a_oH_o$), and decay more slowly:
$|u_m^2|\propto a_o^{2\nu-3}$. In particular the zero-mode,
$\varphi_o$, approaches a constant value at late times.

The general solution for the $y$-dependence of the zero-mode
($m=0$) in a de Sitter cosmology is
\[
\H_o=C_1+C_2{\rm sgn}(y) \coth\mu(y_{\rm h}- |y|)
\left[3-\coth^2\mu(y_{\rm h}-|y|)\right]\,,
\]
where $C_a$ are constants to be determined by the boundary
conditions. The
junction condition in Eq.~(\ref{boundary}) requires $C_2=0$,
and thus
$\H_o=$ constant. The metric perturbation in Eq.~(\ref{pertds}) for the
zero-mode is then given by
\begin{eqnarray}
\label{zeromode}
h_{ij} (t,\vec{x},y)
 &=& a_o^2(t) \A^2(y) E_{ij} (t,\vec{x},y) \,,\nonumber \\
 &=&  C_1\, e^{2H_ot}\, \varphi_o(t)\, 
 \exp({\rm i}\vec{k}\cdot\vec{x})\, \A^2(y)\, 
 \hat{e}_{ij}
\,,
\end{eqnarray}
where $\A(y)$ is given by Eq.~(\ref{Ay}). Thus the bulk dependence of
this mode is the same as that found by Randall and
Sundrum~\cite{RS99} for a Minkowski brane, with $h_{ij}$ decaying away
from the brane proportional to the `warp factor', $\A^2$, which for the
RS brane ($H_o\to0$) is given by $\A^2=e^{-2\mu|y|}$.

Introducing the conformal bulk coordinate $z$, given in
Eq.~(\ref{defz}), and defining $\Psi_m\equiv \A^{3/2}\H_m$, it is
possible to rewrite the off-brane equation in the
Schr\"odinger-like form
\begin{equation}
\label{SE}
 {d^2\Psi_m\over dz^2} - V\Psi_m =-m^2 \Psi_m \,,
\end{equation}
where
\begin{eqnarray}
V(z)&=&{\textstyle{15\over4}}\mu^2\A^2(z)+{\textstyle{9\over4}}H_o^2-
3\mu\left[1+{\rho\over\lambda}\right]\delta(z-z_{\rm b})
\,,\nonumber\\ &=& {15H_o^2 \over 4\sinh^2(H_oz)} +
{\textstyle{9\over4}}H_o^2
- 3\mu\left[1+{\rho\over\lambda}\right] \delta(z-z_{\rm b}) \,.
\end{eqnarray}
In the RS limit, $\rho\to0$ and $H_o\to0$, so that
$\A^2\to[1+\mu(|z|-z_{\rm b})]^{-2}$. The effect in the off-brane
equation of introducing curvature (expansion) on the brane is
two-fold. Firstly it changes the form of $\A(z)$, so that it is no
longer possible to obtain closed form solutions for the Kaluza-Klein (KK) modes
for arbitrary $m$; however, the qualitative behaviour of $\A(z)$
remains the same, i.e., a cusp at the brane, dying off
monotonically to 0 as $z\to\infty$.  The second effect is more
important, and is the introduction of a constant term into $V$, so
that $V>0$ at infinity. This can be treated as a shift in the
effective mass-squared:
\[
m^2\to \hat{m}^2=m^2-m_{\rm c}^2\,,~~ m_{\rm c}=
{\textstyle{3\over2}}H_o\,.
\]
For $m>m_{\rm c}$, the KK modes for a de Sitter brane correspond
qualitatively to those for a flat brane, asymptotically
approaching sinusoidal functions $\Psi_m\sim \exp({\rm
i}\hat{m}|z|)$ as $|z|\to\infty$, but with the reduced effective
mass, $\hat{m}$. For $m< m_{\rm c}$, the effective mass of the
free field at infinity becomes negative, signalling an
instability, and in general these modes will diverge at infinity
and thus be non-normalizable. As remarked earlier, the zero-mode
($m=0$) remains finite (and normalizable) due to the boundary
condition at the brane.

\section{Quantum fluctuations on the brane}

A full treatment of the spectrum of graviton fluctuations
generated by de Sitter inflation on the brane requires a physical
prescription for the boundary condition on the Cauchy horizon at
past null infinity, which in turn requires a quantum cosmological
extension, as considered, for instance, in Refs.~\cite{GS99,HHR}.
However we can gain some insight by simply treating each mode
$\varphi_m$ as a quantum field in four-dimensions with
time-dependent potential, as is done in conventional 4-D inflation
models, where we demand that the fields are in their adiabatic
vacuum state at early-times/small-scales ($k\gg a_oH_o$) well
within the horizon. The extra subtlety in our 5-D model comes from
determining the orthonormal basis for the $y$-dependent functions
and from the need to determine which of these eigenmodes to
include.

As in Ref. \cite{RS99}, we will  require that
the modes be normalizable in the measure $dz$, i.e.,
\begin{equation}
2 \int_{z_{\rm b}}^{z_{\rm r}} |\Psi_m^2| dz = 1\,, 
\label{gennorm}
\end{equation}
where $z$ is the canonical coordinate for the Schr\"odinger-like
Eq.~(\ref{SE}), $z_{\rm b}$ is the position of our brane and $z_{\rm
r}$ is the position of a `regulator brane'~\cite{RS99}.  This
restricts the allowed values of $m$ to the discrete mode $m=0$, and a
discrete spectrum of modes with $m$ such that the boundary condition,
$E'=0$, also holds at $z_{\rm r}$~\cite{RS99}.  As $z_{\rm
r}\to\infty$ this discrete spectrum approaches a continuum of states
for $m> m_{\rm c}$.  The normalization for the continuum modes is
principally determined by the asymptotic sinusoidal form $\Psi_m\sim
\exp({\rm i}\hat{m}|z|)$ as $|z|\to\infty$. However, since the
time-dependent fields $\varphi_m$ are heavy during inflation for
$m>m_{\rm c}$, these modes remain in the vacuum state. Because the
amplitude of these long-wavelength modes decays, their power spectrum
is strongly suppressed on super-horizon scales ($k\ll a_oH_o$).

Normalizing the zero-mode, $\Psi_o=C_1\A^{3/2}$, requires
\begin{equation}
2 \int_{z_{\rm b}}^\infty |\Psi_o^2| dz = 2 \int_{0}^{y_{\rm h}}
C_1^2 \A^2 dy = 1\,, \label{norm}
\end{equation}
which gives
\[
C_1=\sqrt{\mu}\,\,F\!\left({H_o/\mu}\right)\,,
\]
where
\begin{equation}
F\!\left(x\right) =\left\{ \sqrt{1+x^2} - x^2 \ln \left[ {1\over
x}+\sqrt{1+{1\over x^2}} \right] \right\}^{\!\!-1/2}
\!.\label{deff}
\end{equation}
At low energies, i.e., $H_0\ll\mu$ or $\rho\ll\lambda$, we have
$F\approx1$, and recover the RS normalization
$C_1^2\to\mu$. But as the Hubble rate, and thus the energy-scale,
of inflation rises, the value of $|z|$ at the brane increases and
the integral in Eq.~(\ref{norm}) is over a smaller interval, so
that the normalization constant, and hence the amplitude of the
normalized zero-mode at the brane, grows. For $H_o\gg\mu$, or
$\rho\gg\lambda$, we obtain
\[
C_1\approx \sqrt{\mu}\sqrt{3H_o\over2\mu}\gg\sqrt{\mu}\,.
\]

The second-order effective action for tensor perturbations is
\begin{eqnarray}
S_{\rm g}&=&{1\over8\K5^2}\int dt\, d^3\vec{x}\, dy\,na^3
 \nonumber\\ &&~~~\times\left[
{1\over n^2}\dot{E}_{ij}\dot{E}^{ij}-{1\over a^2}\partial_\ell E_{ij}
\partial^\ell E^{ij}-E'_{ij}E^{ij}{}'\right]\,.
\end{eqnarray}
For the zero-mode,
given in Eq.~(\ref{zeromode})
we can integrate over the bulk coordinate $y$,
using the normalization given by Eq.~(\ref{norm}), and this gives 
\[
S_{\rm g}= {1\over8\K5^2}
\, \sum_{+,\times}\, \int d\eta\, d^3\vec{k}\,a_o^2\left[
\left({d\varphi_o \over
d\eta}\right)^2+k^2{\varphi_o}^2\right] \,, 
\]
where we have also integrated over all three-dimensional Fourier modes and
summed the two polarization states.
This has the standard form for a massless graviton in
four-dimensional cosmology~\cite{Grishchuk,Lidsey97}, apart from
the overall factor $1/8\K5^2$ instead of $1/8\kappa_4^2$. It
follows that quantum fluctuations in each 
polarization, $\varphi_o$,
have an amplitude of $2\K5(H_o/2\pi)$ on super-horizon scales. Quantum
fluctuations on the brane at $y=0$, where $\E_o=C_1$, thus have the
typical amplitude
\begin{equation}
\langle {\bar{\varphi}_o}^{~2} \rangle^{1/2}
 = C_1 \langle {\varphi_o}^2 \rangle^{1/2} 
= 2 \kappa_4 \left({H_o\over2\pi}\right)\,
\!F\left({H_o/\mu}\right) \,,
\end{equation}
on super-horizon scales, where $F(H_o/\mu)$ is given by
Eq.~(\ref{deff}).

At low energies ($H_0\ll\mu$), we have $F\approx1$, and we recover
the standard four-dimensional result for gravitational waves
produced during inflation~\cite{Lidsey97}. At high energies
($H_0\gg\mu$), we have $F\approx\sqrt{3H_0/2\mu}$, so that
\begin{equation}
\langle {\bar\varphi_o}^{~2} \rangle^{1/2} \approx
 2 \kappa_4 \left({H_o\over2\pi}\right) \sqrt{3H_o\over2\mu} \,.
\end{equation}
The amplitude of gravitational waves produced during inflation at
high-energies is thus much greater than the standard result in
four-dimensional general relativity.

We are thus able to calculate the primordial spectrum of
gravitational wave perturbations produced by a period of slow-roll
inflation, modelled as quasi-de Sitter expansion on the brane.

\section{Conclusions}

We have discussed the evolution of gravitational wave (tensor-type)
perturbations in a cosmological brane-world scenario. In general the
evolution of the metric perturbations on the brane is coupled to the
evolution in the bulk, and it is not in general possible to separate a
zero-mode corresponding to a massless graviton on the brane, from the
massive states. However we find two special cases in which it is
possible. Firstly on large scales, where spatial gradients on the
brane can be neglected, there is always a zero-mode corresponding to a
homogeneous metric perturbation in the bulk, $E=$const.  Secondly the
equation of motion for the perturbations becomes separable (with the
brane remaining fixed in the bulk) in the case of de Sitter inflation
on the brane.

Modes are characterized by their eigenvalue, $-m^2$, which becomes the
negative mass-squared of the corresponding four-dimensional fields.
We find that there is a discrete normalizable massless mode, $m=0$, as
in the case of a non-expanding brane~\cite{RS99}.  There is also a
continuum of massive KK modes for $m>{3\over2}H_o$, where $H_o$ is the
brane Hubble rate~\cite{GS99}.  This mass gap is a consequence of the
expansion, and may be understood in terms of an effective temperature
of de Sitter spacetime creating a threshold of excitation as suggested
in~\cite{GS99}.

During a period of quasi-de Sitter inflation on the brane, the
continuum of heavy modes remains under-damped and hence these modes
are strongly suppressed on large scales, remaining in their vacuum
state. However the massless mode is over-damped and acquires a
spectrum of classical perturbations on super-horizon scales. By
integrating over the bulk coordinate in the five-dimensional effective
action, we are able to treat the light mode as an effective
four-dimensional field. At low energies, the massless mode is confined
close to the brane, and we recover the standard four-dimensional
result that fluctuations in the massless graviton on large scales are
given by the Hawking temperature, $H_o/2\pi$. At higher energies the
massless mode extends into the five-dimensional bulk and the amplitude
of fluctuations on the brane is enhanced compared with the usual
four-dimensional result.

Inflation at high energy scales thus enhances the amplitude of
super-horizon tensor perturbations that can contribute to CMB
anisotropies on large angular scales. However, the amplitude of scalar
perturbations is also enhanced at high energies, as shown
in~\cite{MWBH99}. While at low energies we recover the standard result
that the ratio of tensor to scalar modes is given by the slow-roll
parameter $\epsilon=M_4^2V'^2/(16\pi V^2)$ (where $V$ is the slow-roll
potential), at high energies this ratio becomes\footnote{This corrects
the high-energy limit of the result given in Eq.~(24) of
Ref.~\cite{MWBH99}, where the amplitude of gravitational waves was
assumed to be the same as the 4-D result, leading to an even smaller
ratio at high energies.}
\[
{A_{\sc t}^2\over A_{\sc s}^2}=6\epsilon
\,\left({\lambda\over\rho}\right)\,,
\]
where $\lambda$ is the constant brane tension and $\rho$ the
inflaton energy density (see \cite{Lidsey97} for the precise definition of 
$A_{\sc t}$ and $A_{\sc s}$). Thus the relative contribution of
gravitational waves to the CMB anisotropies is suppressed for
inflation at energy scales $\rho\gg\lambda$.

Having been stretched beyond the horizon during inflation, the
massless mode remains constant until horizon re-entry during the
subsequent radiation and matter-dominated eras. We have shown that
there is always a zero-mode solution, on sufficiently large scales
($k/a_oH_o\to0$) with $\E_o(y)=$~constant, $\varphi_o=$~constant.
At re-entry the expansion is no longer quasi-de Sitter, and the
massless mode can no longer be decoupled from massive graviton
modes. It remains to be seen whether any significant mixing
between the massless zero-mode and the massive states occurs at
horizon entry. This could lead to very interesting effects if the
lightest continuum modes have a mass of order the Hubble scale (as
is the case during de Sitter expansion), which would lead to a
form of weakly interacting (gravitationally coupled) dark matter
with a very light mass. It is intriguing to note
a recent discussion~\cite{Hu} of the possibility of solving the
problem of excess small-scale structure in the cores of galaxies,
by hypothesizing that the dark matter has a very small mass, of
order the present Hubble scale.


\acknowledgments

We thank Marco Bruni for useful discussions. DW is supported by
the Royal Society. 
DL is grateful to the University of Portsmouth for their hospitality
during his visit, supported by PPARC and CNRS, while this work was
initiated, and DW is grateful to the Observatoire de Paris, Meudon, for
their hospitality during his visit, supported by CNRS, while it was
concluded.



\begin{references}

\bibitem{RS99}
L. Randall and R. Sundrum, Phys. Rev. Lett. {\bf 83},
4690 (1999) [hep-th/9906064].
\bibitem{KIS}
H. Kodama, A. Ishibashi, and O. Seto, hep-th/0004160.
\bibitem{M}
R. Maartens, hep-th/0004166.
\bibitem{Langlois}
D.~Langlois, hep-th/0005025.
\bibitem{BDBL}
C. van de Bruck, M. Dorca, R.H. Brandenberger, and A. Lukas,
hep-th/0005032.
\bibitem{KS}
K. Koyama and J. Soda, hep-th/0005239.
\bibitem{MWBH99}
R. Maartens, D. Wands, B.A. Bassett, and I.P. Heard, Phys. Rev. D,
to appear (2000) [hep-ph/9912464].
\bibitem{BDL}
P.~Bin\'etruy, C.~Deffayet, and D.~Langlois, Nucl.\ Phys.\  {\bf
B565}, 269 (2000) [hep-th/9905012].
\bibitem{BDEL}
P.~Bin\'etruy, C.~Deffayet, U.~Ellwanger, and D.~Langlois, Phys.\
Lett.\  {\bf B477}, 285 (2000) [hep-th/9910219].
\bibitem{SchAdS}
P. Kraus,  JHEP {\bf 9912}, 011 (1999) [hep-th/9910149]; E.E.
Flanagan, S.-H.H. Tye, and I.Wasserman, hep-ph/9910498.
\bibitem{SMS}
T. Shiromizu, K. Maeda, and M. Sasaki, Phys. Rev. D, to appear
(2000) [gr-qc/9910076].
\bibitem{cosmors}
C. Cs\'aki, M. Graesser, C. Kolda, and J. Terning, Phys. Lett.
{\bf B462}, 34 (1999) [hep-ph/9906513]; J.M. Cline, C. Grojean,
and G. Servant, Phys. Rev. Lett. {\bf 83}, 4245 (1999)
[hep-ph/9906523].
\bibitem{MSM}
S. Mukohyama, T. Shiromizu, and K. Maeda, hep-th/9912287.
\bibitem{MFB}
V. F. Mukhanov, H. A. Feldman, and R. H. Brandenberger, Phys. Rep.
{\bf 215}, 203 (1992).
\bibitem{WMLL00} D. Wands, K. A. Malik, D. H. Lyth and A. R. Liddle,
Phys. Rev. D, to appear (2000) [astro-ph/0003278].
\bibitem{bent}
N.~Kaloper, Phys.\ Rev.\ D {\bf 60}, 123506 (1999)
[hep-th/9905210].
\bibitem{GS99}
J.~Garriga and M.~Sasaki, hep-th/9912118.
\bibitem{HHR}
S.~W.~Hawking, T.~Hertog, and H.~S.~Reall, Phys. Rev. D, to appear
(2000) [hep-th/0003052].
\bibitem{Grishchuk}
L.~P.~Grishchuk, Zh. Eksp. Teor. iz. {\bf 67}, 825 -- Sov. Phys.
JETP {\bf 40}, 409 (1974); Ann. N. Y. Acad. Sci. {\bf 302}, 439
(1977).
\bibitem{Lidsey97}
J.~E.~Lidsey, A.~R.~Liddle, E.~W.~Kolb, E.~J.~Copeland,
T.~Barreiro, and M.~Abney, Rev.\ Mod.\ Phys.\ {\bf 69}, 373 (1997)
[astro-ph/9508078].
\bibitem{Hu}
W. Hu, R. Barkana, and A. Gruzinov, astro-ph/0003365.

\end{references}
\end{document}